\documentclass[a4paper,10pt, twocolumn]{article}
\usepackage[utf8]{inputenc}
\usepackage{graphicx}
\usepackage{caption}
\usepackage{listings}
\usepackage{natbib}
\usepackage{makeidx}
\usepackage[breaklinks=true]{hyperref}
\usepackage{breakurl}
\usepackage[table]{xcolor}
\usepackage{color}
\usepackage{authblk}
\usepackage{url}
\usepackage{lmodern}
\definecolor{light-gray}{gray}{0.95}

\title{{PyXNAT}: XNAT in Python}

\author[1]{Yannick Schwartz}
\author[1]{Alexis Barbot}
\author[1]{Benjamin Thyreau}
\author[1]{Vincent Frouin}
\author[2,1]{Gaël Varoquaux}
\author[3]{Aditya Siram}
\author[3]{Daniel S Marcus}
\author[1]{Jean-Baptiste Poline}
\affil[1]{CEA, DSV, I2BM, Neurospin bât 145, Gif-sur-Yvette, France}
\affil[2]{Parietal Team, INRIA Saclay Ile-de-France, Saclay, France}
\affil[3]{Department of Radiology, Washington University School of Medicine, St. Louis, MO, USA}

\definecolor{deep_blue}{rgb}{0,.2,.5}
\definecolor{dark_blue}{rgb}{0,.15,.5}
\definecolor{myblue}{rgb}{.01,0.21,0.71}

\colorlet{deep_color}{deep_blue}
\colorlet{dark_color}{dark_blue}
\colorlet{mycolor}{myblue}

\hypersetup{pdftex,  % needed for pdflatex
  breaklinks=true,  % so long urls are correctly broken across lines
  colorlinks=true,
  urlcolor=mycolor,
  linkcolor=dark_color,
  citecolor=deep_color,
  }

\begin{document}

\date{}
\maketitle

\begin{abstract}

As neuroimaging databases grow in size and complexity, the time researchers spend 
investigating and managing the data increases to the expense of data analysis. 
As a result, investigators rely more and more heavily on scripting using
high-level languages to automate data management and processing tasks. For this,
a structured and programmatic access to the data store is necessary. Web services
are a first step toward this goal. They however lack in functionality and ease
of use because they provide only low level interfaces to databases. We introduce 
here {PyXNAT}, a Python module that interacts with The Extensible Neuroimaging 
Archive Toolkit (XNAT) through native Python calls across multiple operating systems.
The choice of Python enables {PyXNAT} to expose the XNAT Web Services and unify their
features with a higher level and more expressive language. {PyXNAT} provides XNAT users direct
access to all the scientific packages in Python. Finally {PyXNAT} aims to be efficient
and easy to use, both as a backend library to build XNAT clients and as an alternative
frontend from the command line.

\end{abstract}

\section*{Introduction}

The neuroimaging community is producing imaging and related data at an increasing 
rate. Publicly available data and consortia shared data follow the same trend
as funding agencies more routinely require some sharing from the grantees.
The need to share and to maintain data resources at different scales, from large, multi-site studies to
individual laboratories or researchers, has also led to the development of neuroimaging data 
management systems \citep{VanHorn2009}. For instance, the USA-based Biomedical 
Informatics Research Network (BIRN) has during the past ten years developed a 
number of tools to facilitate collaborative research and data sharing in neuroimaging. 
These efforts included the development or use of ontologies \citep{larsonneurolex}, 
data format exchange \citep{gadde2011xcede}, as well as databases and data management systems 
including the Human Imaging Database (HID) \citep{Keator2008}, the LONI IDA 
\citep{VanHorn2009}, and The Extensible Neuroimaging Archive Toolkit (XNAT) 
\citep{marcus2007extensible}. Other consortia and initiatives have also emerged 
to facilitate the handling and sharing of neuroimaging data such as Neurolog 
\citep{Montagnat2008} and CABIG \citep{kakazu2004cancer}. Projects such as the ADNI 
\citep{petersen2010alzheimer} make available high quality large datasets to the community, 
and the number of large multi-modal databases is growing very fast \citep{van2002windows, VanHorn2009}.
Numerous tools for managing the neuroimaging and associated data have been developed as a 
consequence of all these projects. Most of them are built for a specific consortium or 
laboratory, \citep{Ozyurt2010, Montagnat2008}. Fewer are made to be distributed 
as a re-usable stand-alone system. Amongst those, XNAT is one of the most commonly used. It is 
installed in many major institutions\footnote{http://xnat.org/about/xnat-implementations.html}
and enjoys an increasing adoption in the community.

\begin{figure*}[!t]
	\begin{center}
	\scalebox{0.45}{
	\includegraphics{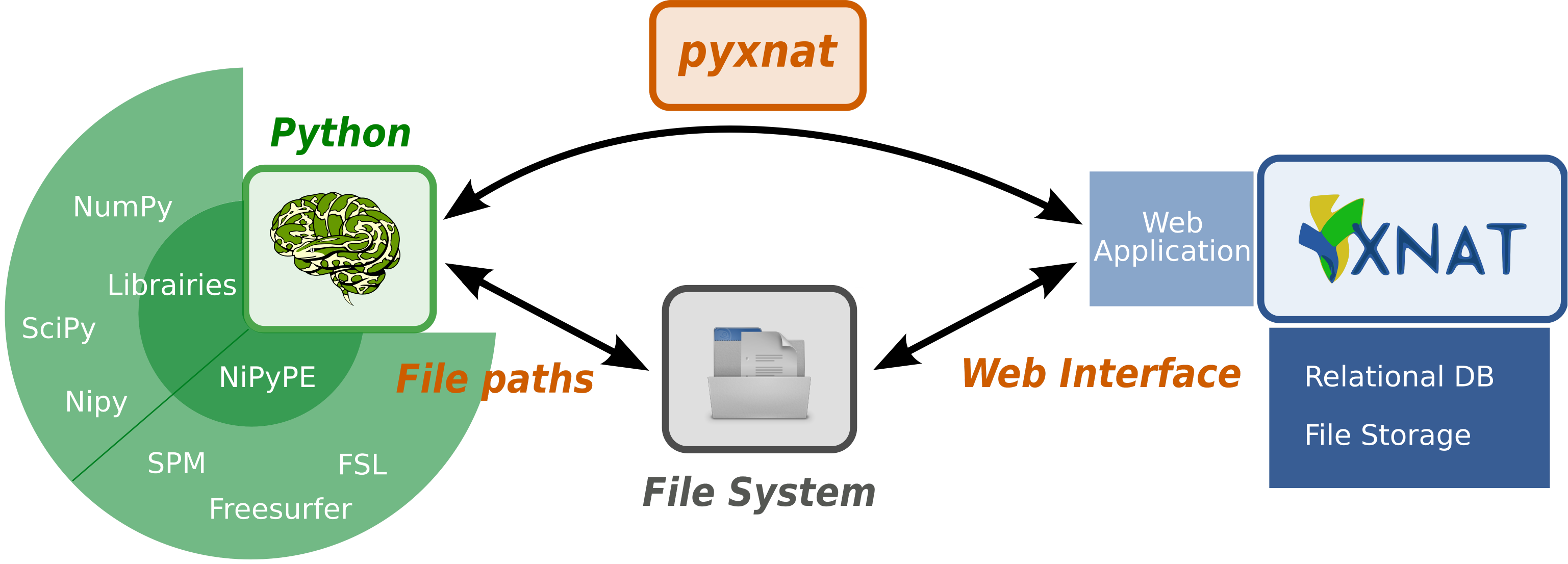}}
	\captionof{figure}{PyXNAT avoids organization on file system.}
	\label{fig:overview}
	\end{center}
\end{figure*}

Databases aim to organize data in a way that it can be efficiently queried and stored. 
There are various database models\footnote{\url{http://en.wikipedia.org/wiki/Database_model}}
\citep{Maier1983, cattell1997object, RENZOANGLES2008} that can be chosen to structure the 
data depending on the problem to solve. 
As the complexity and size of the data increases, neuroimaging databases become harder to use 
because they combine metadata, stored in the database itself, with images, stored in an underlying 
file system. Currently most users manually select and download data through graphical user 
interfaces (GUI), which is intuitive on a small scale, but becomes impractical and error prone 
on a large one. That is because downloading the data on a local disk before processing it breaks 
the way the data is structured in the database. The data has to be organized again, but locally, 
in a consistent layout of files and directories, which effectively duplicates the work done 
setting up the database. Futhermore metadata used to select the data of interest such as 
quality check variables are likely to change during the life span of the database, which makes it even
harder to keep a local dataset synchronized and organized. For example, any manual operations to 
download the data would have to be entirely repeated to reprocess an up-to-date dataset. All
these reasons explain why it is more and more necessary to directly interact programmatically 
with the database. Indeed, relying on a scripted interaction helps maintain consistency 
of accessed data and appropriate versions of the data. This new data flow is represented in
Figure \ref{fig:overview}. In short, large databases 
call for new ways of interacting with and analyzing the data.

\paragraph*{XNAT} Neuroimaging data management systems in general, and XNAT in particular, 
must concurrently solve several issues, depending on specifications of the system. The most
common challenges are: to make the data available in a sustained and secure manner, offer ways 
of searching and querying specific data, and finally to enable updating 
the data repository, either for curation or storing results obtained from processing 
tools.

The XNAT system solved many of these problems. Its core design feature is that 
it models the data through XML schemas, and automatically builds a relational database and a web
interface for accessing the data using that formal description. Many XML schemas are already 
available to describe common neuroimaging or neuropsychological data. These schemas 
greatly ease the work of the data manager constructing the database. XNAT includes many 
useful features such as a permission and access rights system, tabular views, and search 
capacities. Based on this XML description, XNAT users can send queries and receive 
appropriate results. XNAT has moreover developed a representational state transfer Application Program Interface 
(REST API) to push and pull data from an XNAT database. This REST API enables software 
developers and power users to programatically query the server in order to access the data 
and its associated meta. However, REST APIs are low-level interfaces that require 
a significant amount of technical knowledge to perform basic operations.

\paragraph*{PyXNAT}

We developed a Python library ({PyXNAT}) to unify the different REST resources for accessing and 
providing the data to the database as well as ease scripting interactions with an XNAT database. 
Python has recently gained a strong momentum 
in the neuroimaging data analysis community, and more generally in
neuroscience. Combined  
with powerful scientific librairies, it is now close to providing a credible alternative to 
other high level platform independent interpreted language such as MATLAB\texttrademark, but without additional 
licence costs. It offers in addition a very large set of software engineering utilities
such as XML parsing, database, and web interface modules. We discuss more in depth to the 
choice of Python in the next section.

The code of {PyXNAT} was originally developed at Neurospin, (I2BM, CEA, France) 
in the context of the IMAGEN European project\footnote{http://www.imagen.eu} 
\citep{Schumann2010} to help this consortium interact with the IMAGEN database, 
we designed {PyXNAT} to be of general use for the neuroimaging community and licenced it 
under BSD-3\footnote{http://www.opensource.org/licenses/BSD-3-Clause}. The code is 
available\footnote{http://packages.python.org/pyxnat} online and its documentation 
and unit tests coverage quality is kept high so that the programmers external to the 
project can easily contribute. Indeed, {PyXNAT} has started to shift toward a community-based development.

The rest of this article is organized as follows. In the first section, we give some 
background information on the software components on which {PyXNAT} is based. The second
section describes the construction of the library and gives some use case examples. 
Last, we discuss possible limitations and conclude with future improvements.

\section{Material and methods}

We first review the different technologies and components leveraged by {PyXNAT} as a 
preliminary to discussing the implementation design.

\subsection*{Python \label{Python_subsection}} 

We chose to use the Python language since it enjoys a growing success in the 
neuroimaging and neuroscience communities. Indeed, it has recently been subject of a 
special-topic issue in Frontiers in Neuroinformatics entitled "Python in 
neuroscience".  It is a multi-paradigm programming language (for example, 
it supports object-oriented, functional, and procedural programming) with a simple 
and consistent syntax.
It benefits from very efficient open-source scientific packages 
for numerical computation such as NumPy \citep{oliphant2006guide} and  SciPy \citep{jones2001scipy}, 
making it a viable alternative or useful complement to other analysis tools such as 
MATLAB\texttrademark. 
Its flexibility and concise syntax speeds the process of prototyping new algorithms and 
trying out existing softwares. Another strength of the language lies in the variety of its 
application fields, which cover both scientific \citep{langtangen2011primer} and non-scientific
---but relevant---domains such as database management and web development.

Python defines a standard for database interfaces, which is the Python
{DB-API (PEP 249 \footnote{\url{http://www.python.org/dev/peps/pep-0249/}})}.
{PyXNAT} acts as an interface to an XNAT database, but it is a Pythonic wrapping on a REST API
rather than a database driver. As such, it does not really follow a specification based on 
standard database mechanisms and does not replicate operations such as transactions, which
are transparently handled by the XNAT underlying database. However it follows some principles 
from the specification, if not always with the same semantics. As an example, the PEP 249 
defines \texttt{cursor objects} as "[objects] used to manage the context of a fetch operation". 
In other words, these objects are responsible for controlling the data fetching 
but do not do anything when instantiated. They instead rely on 
lazy loading\footnote{http://en.wikipedia.org/wiki/Lazy\_loading} mechanisms that access 
the data only when it is needed. {PyXNAT} design re-uses this principle.

\subsection*{XNAT}

\begin{figure*}[!t]
	\begin{center}
	\scalebox{0.5}{
	\includegraphics{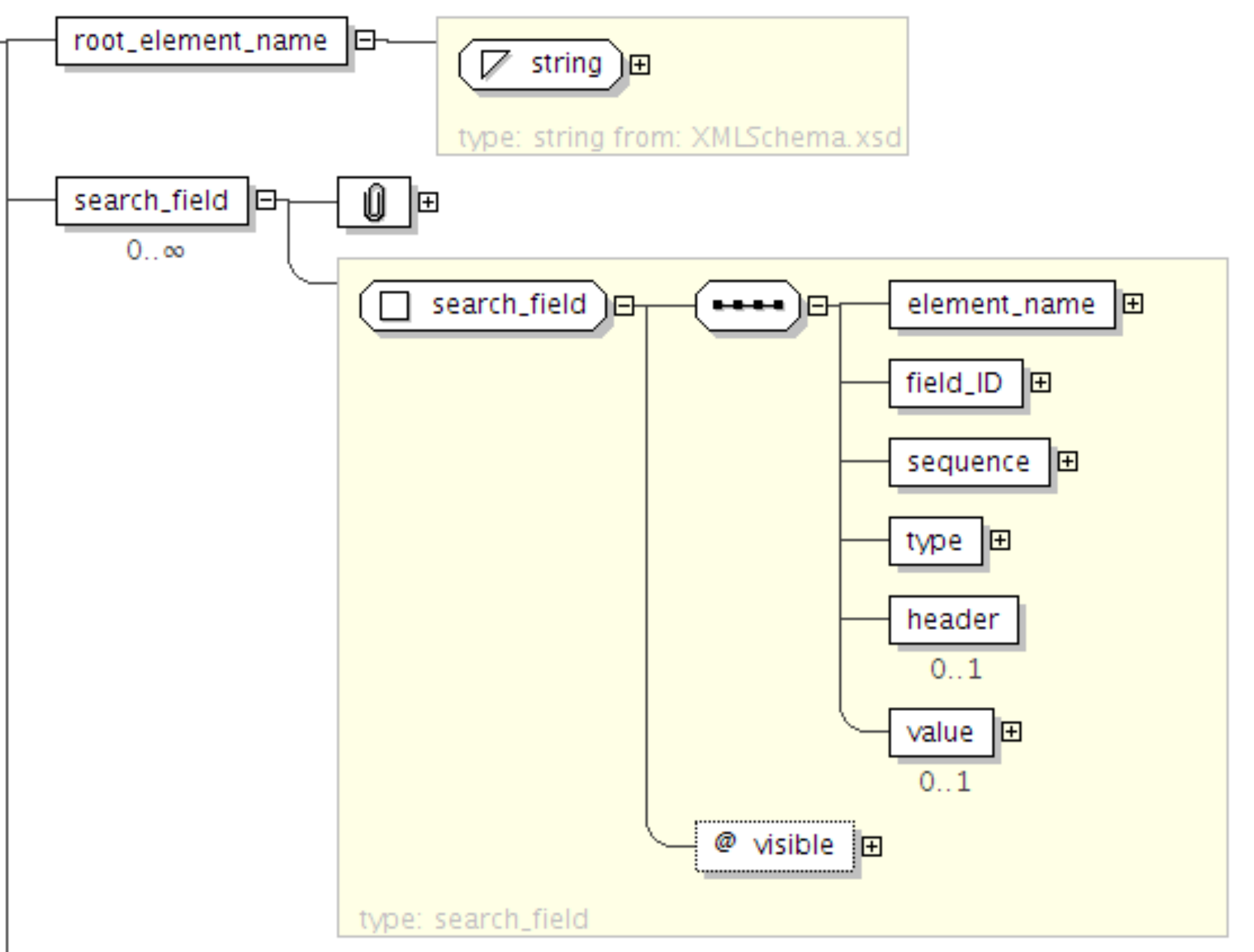}}
	\captionof{figure}{XML Schema for the result table}
	\label{fig:search_engine_table}
	\end{center}
\end{figure*}

\subsubsection*{Overview and key features}

XNAT \citep{marcus2007extensible} is an open source software platform designed 
to manage neuroimaging and related
data. An XML Schema\footnote{http://www.w3.org/XML/Schema} defines the XNAT data model and 
is used to generate a database back-end and a web interface front-end. Using the
XML Schema as an abstraction layer has several advantages. First, the schema 
provides a formal representation of the data in a standard format and 
enables the definition of high-level relations between concepts such as inheritance. 
Second, it makes XNAT extensible since the base schema can accommodate 
extra types for specific studies using custom external schemas.

\subsubsection*{XNAT search engine}

XNAT features a powerful search engine with its own query language that enable users
to search data across all the data types defined in the data model in a transparent 
manner. The query language is specified with an XML Schema document. It enables
standard relational database operations such as projection and selection. The data 
is returned as a CSV or JSON table and can be customized by defining elements in 
the XML query document. The XML schema in Figure~\ref{fig:search_engine_table} specifies
how to format the results as tabular data. {The \texttt{root\_element\_name} 
corresponds to the type of data rendered for each row of the table (e.g., 
{xnat:subjectData}), whereas the \texttt{search\_field} elements defines the columns (e.g., 
{xnat:subjectData/SUBJECT\_ID}). The results of the query are therefore ready to be used 
in any program or spreadsheet software. The XML Schema in 
Figure~\ref{fig:search_engine_criteria} defines how to express search predicates for XNAT.
The main element of the query is the \texttt{criteria\_set} element, which can nested with
\texttt{child\_set} elements in order to perform more complex queries (e.g., return subjects
that are over 20 years old and left-handed, or subjects that are under 20 and right-handed).
The \texttt{criteria\_set} element takes a \texttt{method} value which indicated which
boolean operator to use (AND or OR). Each \texttt{criteria\_set} is composed of a list
of constraints, defined by a \texttt{schema\_field} (e.g., xnat:subjectData/HANDEDNESS), a
comparison operator, and a \texttt{value}. The query language therefore enables usual
operations such as criteria comparison and nesting.

To promote interaction between different users of the same database and help system
administrators, the XNAT search engine provides a way to share queries for all or a 
subset of users. 

\begin{figure*}[!t]
	\begin{center}
	\scalebox{0.4}{
	\includegraphics{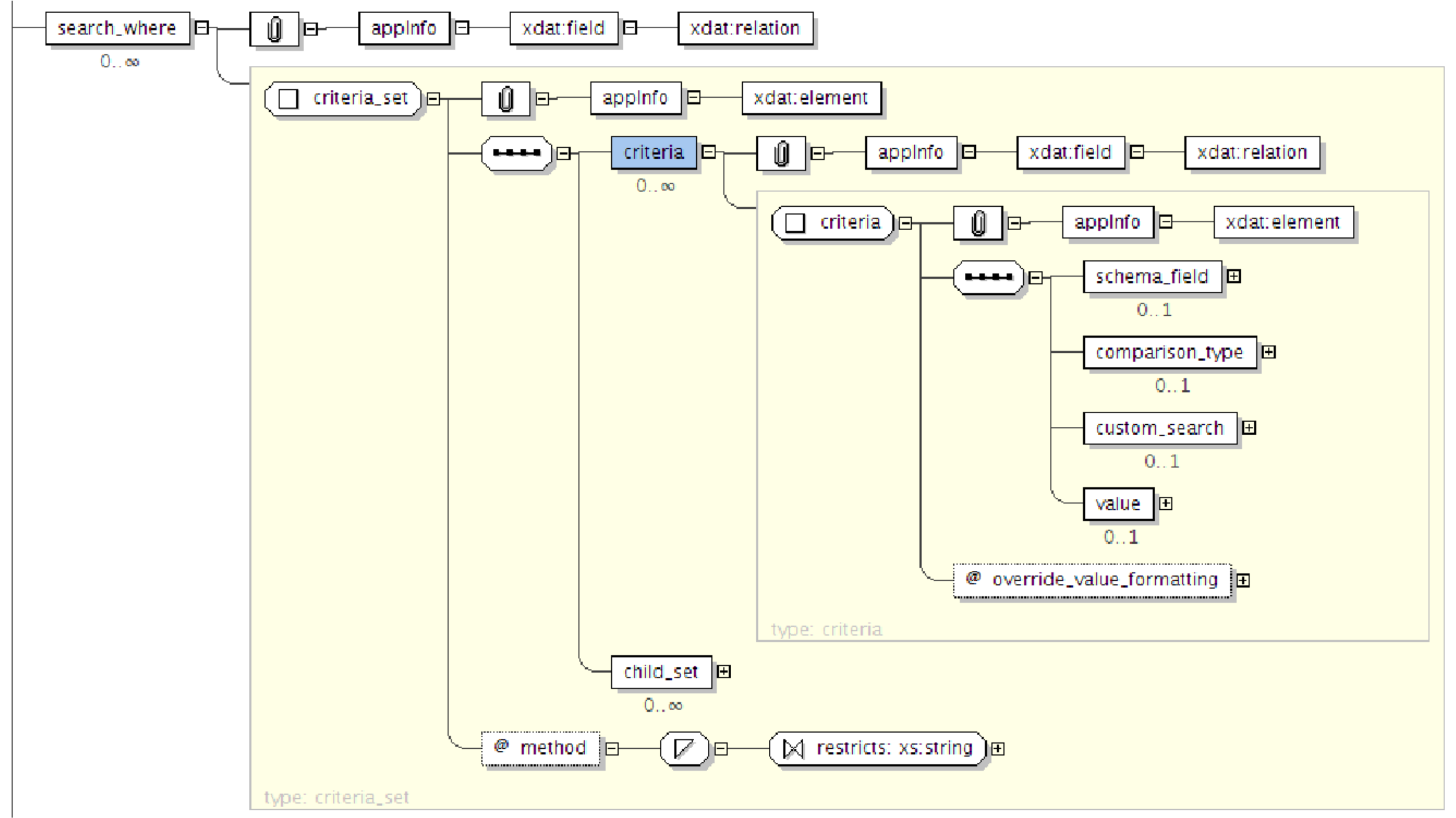}}
	\captionof{figure}{XML Schema for the search criteria}
	\label{fig:search_engine_criteria}
	\end{center}
\end{figure*}

\subsubsection*{The REST model}

REST (Representational State Transfer) \citep{fieldingrepresentational} is a generalization of the
architectural principles of the World Wide Web and is used to develop web services 
alongside or as an alternative to other 
specifications such as the Simple Object Access Protocol or the Common 
Object Request Broker Architecture. A RESTful architecture identifies a 
set of resources, which can be entities or collections, with standardized 
Uniform Resource Identifiers (URIs). The methods to interact with the resources rely
on HTTP verbs---such as GET, PUT, POST and DELETE---that are mapped to 
resource-specific semantics. This means that resources map to a set
of views to represent the data state on the server independently from the way it is stored.
REST also allows representing the resources content in different formats (e.g.,
XML, HTML, and plain text). 

XNAT uses URIs' generic syntax, which consists in a sequence of component parts
describing the communication protocol, the resource location, and additional
information. An example inspired from the RFC 3986\footnote{http://www.ietf.org/rfc/rfc3986.txt}
summarizes the syntax:

\footnotesize
\begin{center}
\begin{samepage}
\begin{verbatim}
 http://central.xnat.org/REST/projects?format=csv
 \__/   \______________/ \___________/  \_______/
  |           |               |             |
scheme    authority         path         query
\end{verbatim}
\end{samepage}
\end{center}
\normalsize

RESTful architectures organize resources in a hierarchy. Basically, URIs' paths
are constructed using a fixed set of keywords that have parent-child relations. In 
XNAT, the main concepts follow the tree structure represented in 
Figure~\ref{fig:XNAT_REST_model}. Keywords are paired with an ID to point to
a specific resource. Collection resources return a list of identifiers and do not end 
with an ID. Table~\ref{tbl:REST_resources} illustrates how XNAT uses the REST resources
to list the project names on a server and access a specific one.

\footnotesize
\begin{center}
    \begin{tabular}{ll}
        \hline
        \\
    	\textbf{Resource type} & \textbf{Path}\\
    	element resource & /REST/projects\textbf{/PROJID}\\
    	collection resource & /REST/projects\\
    	\\
    	\hline
    \end{tabular}
    \captionof{table}{XNAT URI design}
    \label{tbl:REST_resources}
\end{center}
\normalsize

\begin{figure}[h]
	\begin{center}
	\scalebox{0.03}{
	\includegraphics{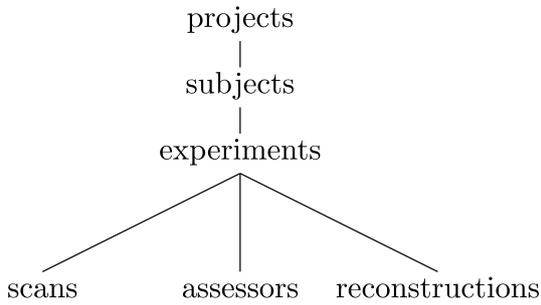}}
	\captionof{figure}{XNAT REST model}
	\label{fig:XNAT_REST_model}
	\end{center}
\end{figure}

URIs support a range of operations through the HTTP verbs. Collection 
resources typically only support the GET method whereas element resources
use GET, PUT and DELETE to support access, creation, and deletion operations.
To perform additional operations, XNAT leverages the query component of URIs.
As shown in Table~\ref{tbl:XNAT_query_string}, it enables selecting and filtering 
the outputs as well as choosing the output format.

\footnotesize
\begin{center}
    \begin{tabular}{ll}
       	\hline
    	\\
    	\textbf{Option} 	& \textbf{URI query string}		\\
		select output 		& ?columns=ID,project			\\	
    	filter output 		& ?xsiType=xnat:mrSessionData	\\	
    	output format 		& ?format=csv					\\
		\\
    	\hline
    \end{tabular}
    \captionof{table}{URI query strings usage in XNAT}
    \label{tbl:XNAT_query_string}
\end{center}
\normalsize

The XNAT REST API is separated in two parts: the hierarchical structure described on 
Figure \ref{fig:XNAT_REST_model} and the search engine. Navigation through the database 
and downloading files attached to subjects or experiments is accessible from the URIs 
whereas getting tables containing metadata is enabled by the search engine.

\section{Results}

We implemented the {PyXNAT} package on top of the XNAT REST API to enable easy 
communications with XNAT through the Python language. In this section, we 
describe the general design of the library as well as specific mechanisms that
are original or of particular importance. We finish by giving some examples of real 
life uses cases for {PyXNAT}.

\subsection*{Architecture and design} 

{PyXNAT} combines several components to interact with an XNAT server, which are described
in Figure \ref{fig:pyxnat_architecture}. Its core 
relies on the httplib2 Python module, which is in charge of issuing calls to
XNAT. The REST structure itself, which is described Figure~\ref{fig:XNAT_REST_model}
is static and cannot be discovered with the XNAT REST API. This is why {PyXNAT} 
uses a configuration file to model the REST structure and to generate a programming 
interface that maps the REST API to Python objects and methods. The modelling of the
REST API is used to generate HTTP calls against XNAT as well as parsing the 
responses to generate Python objects. 

\begin{figure}[b]
	\begin{center}
	\scalebox{0.80}{
	\includegraphics{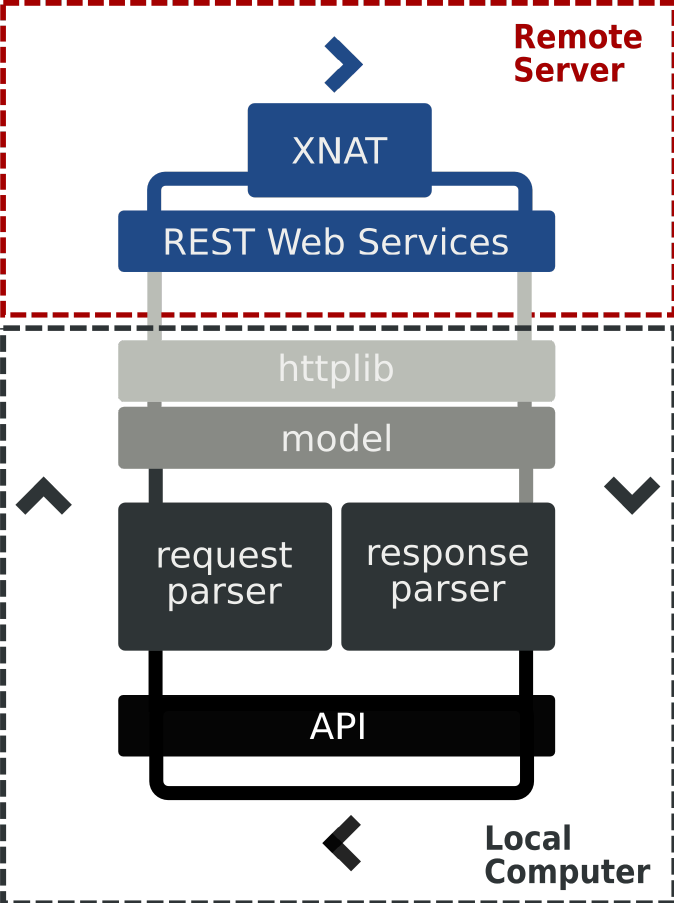}}
	\captionof{figure}{PyXNAT architecture}
	\label{fig:pyxnat_architecture}
	\end{center}
\end{figure}

The XNAT REST API is composed of a set of hierarchical resources, which is the 
REST structure itself, and an endpoint for its search engine. The REST structure 
gives access to files (e.g., images), as well as metadata. The URIs look a 
lot like files and directories paths on a file system and 
can effectively be viewed as such. The search engine gives access to the metadata and 
enables searching them, but does not provide any mechanism to point to files. So with the REST 
API, it is possible to look for all the subjects that are 14 years of age or have a 
specific answer to an assessment. But the search will not be able to yield URIs to files. 
{PyXNAT} builds on XNAT by using the results of the search engine to subsequently generate URIs and 
retrieve files. By tightly integrating those two mechanisms, {PyXNAT} delivers a more powerful and 
succint way to interact with XNAT. For example, 
{PyXNAT} is able to retrieve all the assessments from a subject in a single statement whereas the 
REST API from XNAT would require several calls.

\subsubsection*{Object Mapper}

{PyXNAT} borrows language elements from SQL (Structured Query Language) 
and the Python DB-API to define a familiar and easy to use query language. 
The \texttt{select} 
statement defines the data to return from a query, either a list of 
identifiers if it is a \texttt{collection object}, or a Python object pointing to a 
single URI if it is an \texttt{element object}. An object mapper
returns Python objects reflecting URIs, and offering actions through 
their methods. These actions include, resource operations, database browsing,
and files downloading and uploading.

\footnotesize
\begin{lstlisting}[language=Python]
interface.select('/projects')
interface.select('/project/PROJID')
\end{lstlisting}
\normalsize

The element objects share common operations for insertion and deletion for 
example, but also feature specific methods. For example all types of entities
can be created and deleted, but only the project objects may handle the access
permissions:

\footnotesize
\begin{lstlisting}
p = interface.select('/project/PROJID')

p.insert()
p.delete()
p.exists()
p.set_accessibility('public')
\end{lstlisting}
\normalsize

\texttt{Collection objects} use a lazy loading mechanism and work essentially the same 
way as \texttt{cursor objects} as defined in Python DB-API. The object itself doesn't
issue any request on the database and delegates the actual query to dedicated 
methods. The table~\ref{tbl:collection_methods} summarizes and compares the 
collection or cursor objects methods.

\footnotesize
\begin{center}
    \begin{tabular}{ll}
	    \hline
	    \\
    	\textbf{Python DB-API} & \textbf{pyxnat-API}\\
    	fetchone() & first() or fetchone()\\
    	fetchmany() & not-supported\\
    	fetchall() & get() or fetchall()\\
    	\\
    	\hline
    \end{tabular}
    \captionof{table}{\texttt{Cursor objects} methods comparison.}
    \label{tbl:collection_methods}
\end{center}
\normalsize

{PyXNAT} container objects are implemented as Python generators. Generators provide
a convenient mechanism for lazily looping over items (e.g., using a \texttt{for} loop) without 
accessing or loading those items until they are needed. For example, to perform an 
operation on every subject from all projects, {PyXNAT} operates on the
subjects from each project one at a time.
Without the lazy access mechanism, the library would have to retrieve all the 
subjects from all the projects before operating on them. This might take a 
considerable amount of time that can be used instead to start operations 
that needed the subjects in the first place. While there are other ways
to iterate over subjects, this example demonstrates the flexibility introduced by the 
{PyXNAT} library:

\footnotesize
\begin{lstlisting}[language=Python]
s = interface.select('/projects/*/subjects')

for subject in s:
	<perform operation>
\end{lstlisting}
\normalsize

The keywords used in the \texttt{select} statement are the same as the ones defining
the REST structure represented in Figure~\ref{fig:XNAT_REST_model}. The ability
to chain those keywords through the Python objects enable users to
express more complex queries very easily. For example, the following calls, which
are equivalent, return all the files for all the experiments related to 
a subject in any project in the database. Of course, it would be as easy to use
identifiers or more constrained filters instead of the wildcard "\texttt{*}" 
in order to return a specific set of files. The different syntaxes all rely on the 
same underlying Python objects. They exist because there are two different ways to
use {PyXNAT}. First, as a library, which calls for efficiency and enables to reference
directly the data. Second, as an interactive command line frontend for XNAT, in which
case performance is not the main concern. The need to quickly explore the database 
is, however, far more important and explains the introduction of shortcuts in the syntax.

\footnotesize
\begin{lstlisting}[language=Python]
interface.select.projects().subjects().experiments().resources().files()

interface.select('/projects/*/subjects/*/experiments/*/resources/*/files')

interface.select('//experiments//files')
\end{lstlisting}
\normalsize

\begin{figure*}[!t]
	\begin{center}
	\scalebox{0.8}{
	\includegraphics{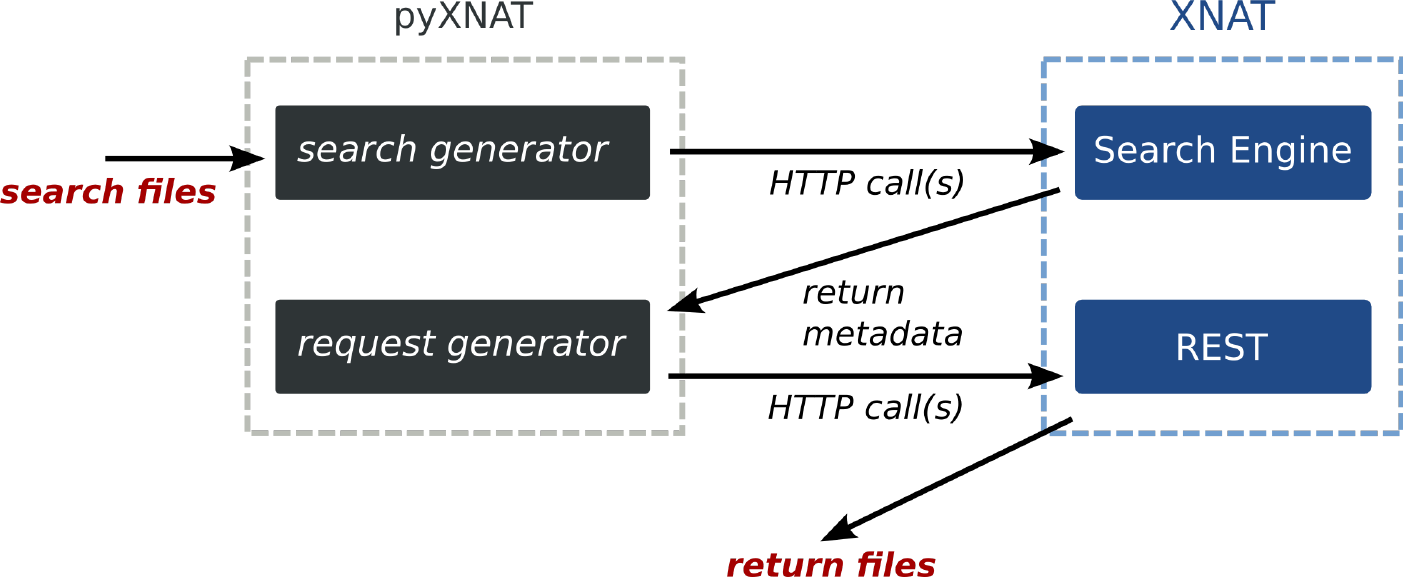}}
	\captionof{figure}{PyXNAT integration of the search engine and the files access}
	\label{fig:pyxnat_ws_integration}
	\end{center}
\end{figure*}

\subsubsection*{Search integration}
\label{section:search_integration}

XNAT's search engine sets up queries using the dataypes defined in the
XML schemas. It is accessible from a single URI, on which it is possible to 
POST---basically send---an XML file describing the query. The request gets
results in form of a CSV (comma-separated values) table, which contains the
requested data and identifiers. However, the URIs referencing files cannot be
returned because they are not stored in the database. Moreover, the 
REST API does not provide any mechanism to use the identifiers to build URIs
referencing files. {PyXNAT} deals with the complexity of writing the XML documents 
and offers a simple language to use the search engine. It also parses the outputs
from the search engine to generate valid URIs that get resources on XNAT. Thoses
mechanisms are illustrated in Figure \ref{fig:pyxnat_ws_integration}.

To keep the semantics consistent throughout the API, {PyXNAT} uses again its \texttt{select}
statement to define the data to capture, but with different parameters. The first
argument is specifies the type of an entry, and the second
argument is a list of fields and defines the columns of the table to return. The SQL
\texttt{where} clause is replicated in {PyXNAT} to formulate the search criteria.
The criteria set is expressed as a list of tuples, where each tuple corresponds to 
a single search constraint. A constraint tuple is a 3-value entity composed of a 
search field, a comparison operator and a value. Every query may include sub-queries, 
expressed by lists of tuples. The whole request with {PyXNAT} is therefore expressed 
with an SQL-like syntax which is close to other query languages and enables to fully 
leverage the XNAT search engine. This syntax is close to SQLAlchemy's\footnote{http://www.sqlalchemy.org}, 
a popular Python ORM library (Object-relational Mapping)\footnote{\url{http://en.wikipedia.org/wiki/Object-relational_mapping}}.

\footnotesize
\begin{lstlisting}[language=Python]
# PyXNAT search example
row = 'xnat:mrSessionData'
columns = [
	'xnat:subjectData/LABEL', 
	'xnat:mrSessionData/AGE', 
	'xnat:subjectData/GENDER'
]
	
criteria = [
	('xnat:mrSessionData/PROJECT', 
		'=', 'MY_PROJECT'),
	('xnat:mrSessionData/PROJECT', 
		'=', 'CENTRAL_OASIS_CS'),
	'OR'
]
	
interface.select(row, columns).where(criteria)

# SQLAlchemy example from the online documentation
session.query(User).filter(User.name.like('%ed'))

\end{lstlisting}
\normalsize

The same syntax can be used to combine the search engine with the hierarchical REST 
resources. The \texttt{select} statement from the object mapper can be chained with 
the \texttt{where} clause, which uses the search engine. {PyXNAT} returns objects, for 
which subjects match the criteria defined in the \texttt{where} clause.

\footnotesize
\begin{lstlisting}[language=Python]
interface.select('//experiments').where(criteria)
\end{lstlisting}
\normalsize

\subsection*{Database introspection}

XNAT databases contain several kinds of data such as imaging data, demographic information,
behavioral data, and experimental design. These categories are further differentiated; 
for instance, imaging data may be acquired from several different modalities including 
structural and functional MRI as well as PET. Further complicating the situation, these 
image modalities are often referred to by several different names. For example, a database
may reference a T1-weighted image as just "T1", whereas another one reference it as "MPRAGE".
This terminology problem can be adressed with ontologies and data integration technologies
which are currently being developed by the community \citep{bug2008nifstd, larsonneurolex}. 
However as a first step adressing this problem, {PyXNAT} provides functions to retrieve list
of all values used in a given database. This functionality gives users the ability to
interact with the data and the data model, in order to quickly provide a summary of a large
number of data types and entries. 

XNAT provides basic introspection methods that are replicated and augmented in {PyXNAT}. In 
particular, XNAT provides REST functions to query the data types that are defined by the XML 
Schema. These functions enable users to learn from a database, for example, that it defines
the concept of subject and that a subject has a gender or tha the \textit{age} of the subject
is actually defined for an \textit{experiment} performed on this subject. However, XNAT lacks
a helper function to extract all the values that a data field takes in a specific database.
To provide this functionality with a consistent API, {PyXNAT} uses the search engine from
XNAT. This is very useful when building queries, since it provides a list of all values that
can be used. In the interactive session below, we show the
different methods that enables users to explore the data model and find data:

\footnotesize
\begin{lstlisting}[language=Python]
# retrieve list of datatypes
>>> interface.inspect.datatypes()
[..., 'xnat:mrSessionData', ...]
# retrieve list of datatypes fields
>>> interface.inspect.datatypes('xnat:mrSessionData')
[..., 'xnat:mrSessionData/AGE', ...]
# retrieve list of field values
>>> interface.inspect.field_values('xnat:mrSessionData/AGE')
[..., '14', '25', '42',...]
\end{lstlisting}
\normalsize

\subsubsection*{Cache}

{PyXNAT} maintains a local copy of all the server responses into a cache (i.e., the
data files that may be images as well as the metadata arrays or resources 
listing). The main goal of the cache is to improve performance; but, it can also 
be rendered persistent and provide a full "offline" mode to {PyXNAT}.

The {PyXNAT} cache is primarily an implementation of the HTTP caching mechanism
that stores the data on a filesystem. HTTP is known as a request-response 
protocol, which means that a client sends a request to a server that is 
responsible for processing and returning a message in response. The message
is composed of two main parts: the header and the body. The header contains
information on the message and on the server. The body contains the actual data
that was requested (e.g., an image). HTTP provides cache validators in the
header of the messages to transmit the status of the resource to the client,
which can take a decision on the validity of the cached version of the resource. 
This strategy prevents the client from downloading unnecessary data and improves 
performance by reducing the network traffic. XNAT only provides the 
"Last-Modified" field in the header, which can be checked against the date of 
the local version of the data. Only the resources that link to a file support 
the cache validation in XNAT. The other resources --- elements listing, metadata 
values --- need to be downloaded again to make sure the local data is up to 
date. This is why {PyXNAT} introduces an additional expiration mechanism to avoid 
repeatedly requesting resources to the server for certain operations. In 
other words, if a cached resource is accessed within a specified amount of time 
(default to one second), the data will not be downloaded again.

% XXX: Gael: Do you have any cache replacement policy? If so, what is it?
% If not, you should point this out as a limitation.

\subsection*{Database management}

{PyXNAT} supports additional XNAT functionalities including user, project, and
pipeline management as well as search utilities. Those features reflect exactly what is
provided by the XNAT REST API. We now present two critical interfaces. 

The first interface provides project management functionality. 
It enables project owners to configure their project, add users, and set up access permissions. 
This is achieved mainly by configuring two attributes: the user role and the project 
accessibility.

\footnotesize
\begin{lstlisting}[language=Python]
interface.manage.project('ID').set_accessibility(level)
interface.manage.project('ID').add_user('user', role)
\end{lstlisting}
\normalsize

The second interface is the search utility. It enables users to create and share
searches with other users. It uses the same syntax as the one described for the \texttt{where}
clause. An additional {PyXNAT}-specific feature is the ability to create search
templates. These templates maintain the ability to be shared between users, but instead of carrying
values, they define keys to be replaced by the actual value when used. This makes it 
possible to easily re-use any kind of search.

\footnotesize
\begin{lstlisting}[language=Python]
criteria = [('xnat:subjectData/GENDER', '=', 'male'), 'AND']
interface.manage.search.save('search_name', row, columns, criteria, users)
interface.manage.search.get('search_name')

criteria = [('xnat:subjectData/GENDER', '=', 'gender'), 'AND']
interface.manage.search.save_template('template_name', row, columns, criteria, users)
interface.manage.search.use_template('template_name', {'gender':'male'})
\end{lstlisting}
\normalsize

\subsection*{Usage examples}

{PyXNAT} is a powerful and easy to use library to build client applications for XNAT.
As an example, NiPyPE \citep{Ghosh2010} is a Python module that interfaces to existing
neuroimaging software such as SPM, FSL or FreeSurfer. It is also able to distribute
jobs over clusters, which makes it very efficient to process large amounts of data. 
Its data connection method was originally file system based but it can now in addition 
access an XNAT server through {PyXNAT}. {PyXNAT} and NiPype are being used jointly to run 
analysis on IMAGEN, which is a European project that aims to study addiction risk factor 
in a database containing over 2000 adolescents. Figure~\ref{fig:pyxnat_nipype} depicts 
how the two packages interact.

\begin{figure*}[!t]
\begin{center}
\scalebox{0.50}{
\includegraphics{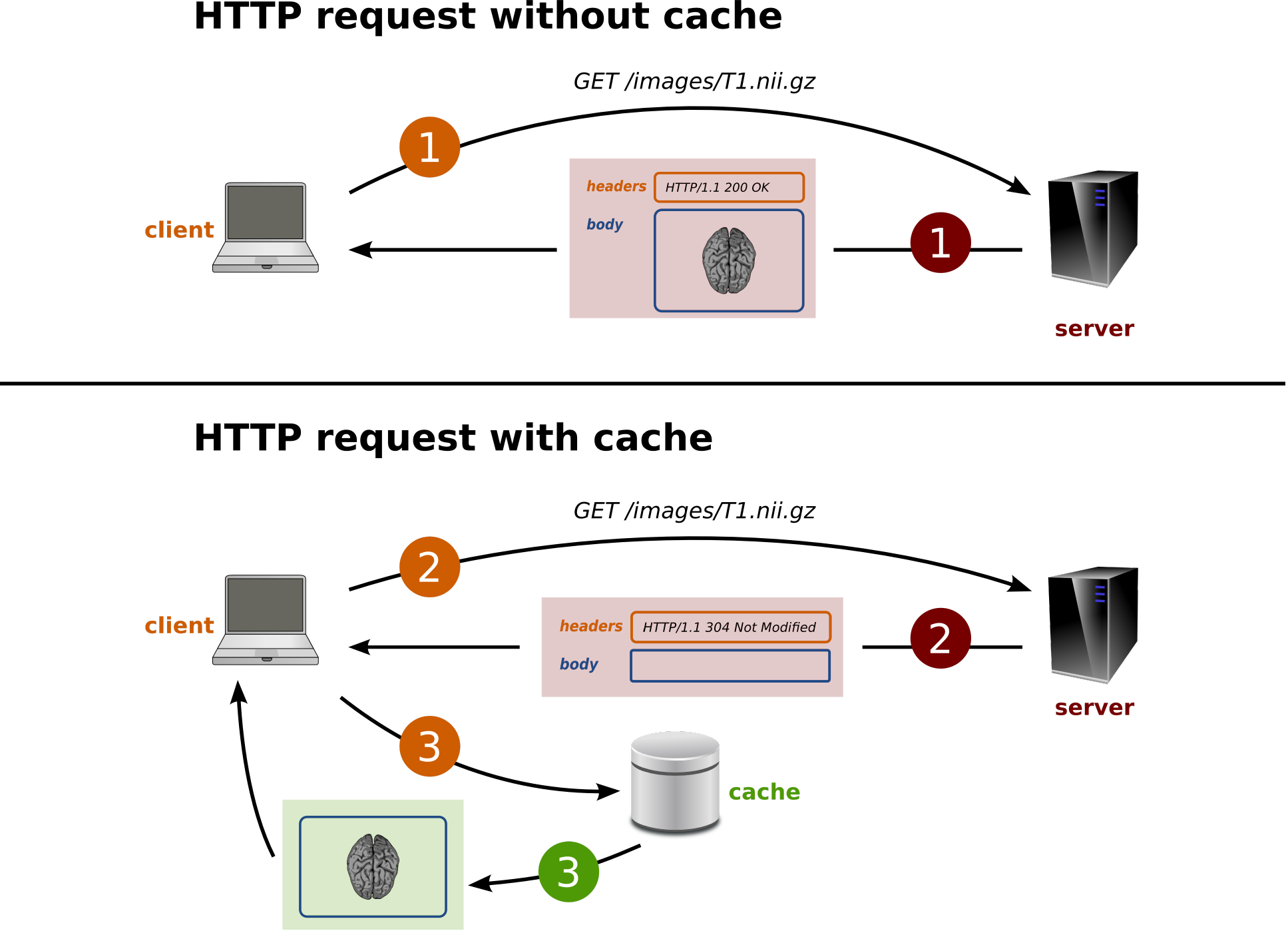}}
\captionof{figure}{HTTP cache mechanism}
\label{fig:httpcache}
\end{center}
\end{figure*}

\begin{figure}[p]
	\begin{center}
	\scalebox{0.55}{
	\includegraphics{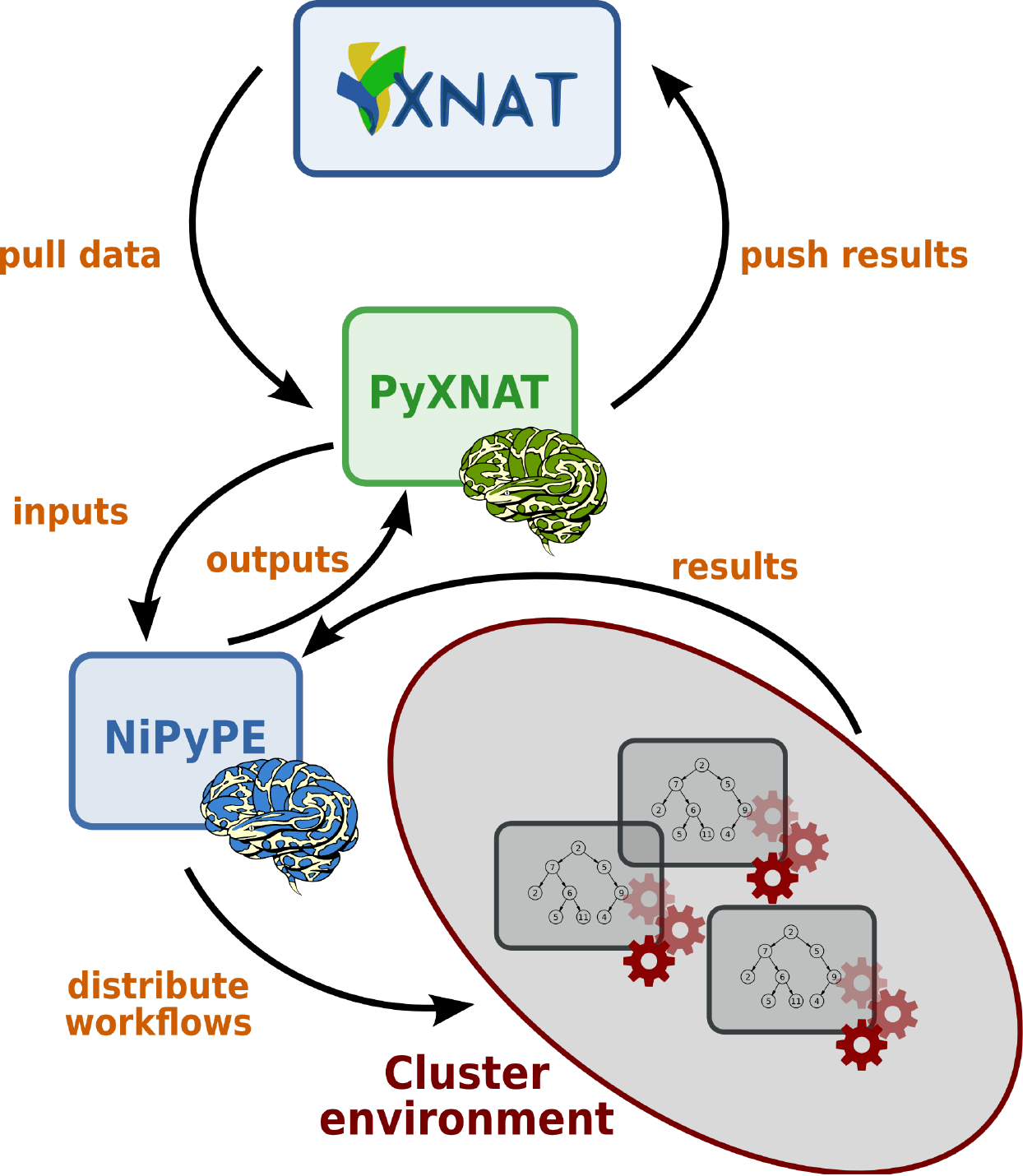}}
	\captionof{figure}{PyXNAT and NiPyPE interactions}
	\label{fig:pyxnat_nipype}
	\end{center}
\end{figure}

Other projects have started or already support XNAT through {PyXNAT}. Among them 
are the XNAT tools from the XNAT group, which were originally written in Java  
and are currently being re-written in Python using {PyXNAT}. 
Another example is the Connectome Viewer \citep{Gerhard2011a}, which can now read 
and write data on XNAT.

\section{Discussion and Conclusion}

Historically, neuroimaging researchers have used ad-hoc procedures for maintaining
their analysis and data. Over the last few years, there have been several attempts to
build databases to manage neuroimaging data. The ability to programmatically access 
neuroimaging databases is becoming increasingly important to perform batch analysis 
and administration tasks, as their growth makes them all but impossible to operate manually. 
However it has been difficult to use standard analysis tools and these database systems together.

One of the most widely used neuroimaging database systems is XNAT. XNAT is an open-source 
database that incorporates many useful and powerful features including an efficient search 
engine and a REST API. However, being written in Java and having a REST API, it offers no
natural bridge to the most common analysis tools, that are accessible from a scripting language 
(like Matlab or Python) or from the command line.

{PyXNAT} provides a bridge between XNAT and analysis tools. Combined with an interactive
Python terminal such as IPython \citep{perez2007ipython}, it can also be used as an 
alternative front-end for XNAT. Since it is written in Python, it becomes readily accessible 
to the vast and relevant set of Python tools in the neuroscience domain. 
Moreover, command
line tools can easily be developed using PyXNAT, as such, popular programming languages can 
easily benefit from PyXNAT merely by issuing system calls. The XNAT team is currently
rewriting its command line tools using PyXNAT. Most programming languages provide bridges
toward other languages. As an example, PyMat 
\footnote{http://claymore.engineer.gvsu.edu/~steriana/Python/pymat.html} enables
Python scripts to execute Matlab commands, and pass back and forth
Numpy arrays. While these solutions are not as robust as keeping
a single programming environment, they provide a viable option
to use existing code. Another solution to use PyXNAT from other languages
is to use the softwares wrapped by NiPyPE.

{PyXNAT} focuses on ease of use, combining RESTful services with 
clear semantics and adding helper features. It also makes {PyXNAT} highly efficient: being 
a thin layer over HTTP with a cache mechanism, it is at least as efficient as native REST 
calls. The package is open-source under the BSD license. It is available for 
download\footnote{http://pypi.python.org/pypi/pyxnat/} and has an online 
documentation\footnote{http://packages.python.org/pyxnat/}, which covers installation and usage.

There are several areas where {PyXNAT}, in its current version (0.9.3), needs to be improved. 
One of the most important areas that
could be improved is the {PyXNAT} cache, which is currently only disk-based.
If several processes share the same cache folder, one has to be careful 
to avoid concurrent read and write operations on the same files. The cache
could be replaced by a full featured local database. It would support concurrent access
and also enable an offline mode for {PyXNAT} with search capabilities.
One could also add synchronization features to update the local database and push
back generated results to the remote server. Users would then be able to work 
seemlessly on and offline. Other possible improvements include a logging framework to trace all 
the REST calls, advanced data filtering capabilities from the REST API, or the prearchive
mechanism from XNAT.

{PyXNAT} could be used to develop a federation layer between XNAT servers. 
It would mostly help to access the data, but using its introspection
functions, it could issue simple queries on multiple XNAT instances. 
PyXNAT could also 
help to federate heterogeneous databases systems, but as a component along similar 
librairies. The complex challenges of data integration, such as data alignment would 
however have to be addressed separately. The INCF (International Neuroinformatics 
Coordinating Facility), and in particular its datasharing task force, is currently working 
on these issues. For example, the datasharing task force is working on an API for accessing
different neuroimaging databases (XNAT, HID, IDA, ...), that could eventually be re-used in 
PyXNAT. Its goal is not to promote a specific database, but rather standards and 
methods to share and re-use neuroimaging data. However due to its popularity and relevance,
XNAT and therefore {PyXNAT} are part of the components being used to build prototypes for 
the initiative.

In conlusion, {PyXNAT} enables XNAT access in the Python environment. It can be used 
both as an interactive command line interface and as a back-end communication library. 
We see {PyXNAT} as an major step to help process and administrate datasets in XNAT servers.

\section{Acknowledgements}

We thank Jarrod Millman for helpful reading of the original manuscript. We also thank the 
NIPY community for their tools and advice in general, and all the {PyXNAT} users for their 
feedback and patience. Support was provided by the IMAGEN project, which receives research 
funding from the European Community's Sixth Framework Programme (LSHM-CT-2007-037286). 
This manuscript reflects only the author's views and the Community is not liable for any 
use that may be made of the information contained therein.

\bibliographystyle{chicago}
\bibliography{biblio.bib}

\onecolumn
\section{Appendix}

This small example illustrates how to download T1 images from subjects over 80 years old
on XNAT Central\footnote{https://central.xnat.org} with {PyXNAT}, and process them in 
parallel on a computer. For the sake of simplicity, we chose the BET command line tool,
which extracts the brain from the image of the whole head, as an analysis example. 
The functions, that distribute the processing on several processors, are part of 
the standard library of Python. The example is also available on 
github\footnote{https://gist.github.com/1816347}, and requires 
FSL\footnote{fsl-bet path may have to be changed in the script to match your installation} 
and pyxnat version 0.9.3 or above to run. The script will prompt the user for a login and
password, so one may need to first register on XNAT CENTRAL.
\\

\footnotesize
\begin{lstlisting}[language=Python]

import os
from subprocess import Popen
import multiprocessing as mp

import pyxnat

URL = 'https://central.xnat.org' # central URL
BET = 'fsl4.1-bet2'              # BET executable path

central = pyxnat.Interface(URL) # connection object

def bet(in_img, in_hdr): # Python wrapper on FSL BET, essentially a system call
    in_image = in_img.get()     # download .img
    in_hdr.get()                # download .hdr
    path, name = os.path.split(in_image)
    in_image = os.path.join(path, name.rsplit('.')[0])
    out_image = os.path.join(path, name.rsplit('.')[0] + '_brain')
    print '==> %s' % in_image[-120:]
    Popen('%s %s %s' % (BET, in_image, out_image), 
          shell=True).communicate()
    return out_image

notify = lambda m: sys.stdout.write('<== %s\n' % m[-120:]) # print finish message
pool = mp.Pool(processes=mp.cpu_count() * 2) # pool of concurrent workers
images = {}
query = ('/projects/CENTRAL_OASIS_CS/subjects/*'
          '/experiments/*_MR1/scans/mpr-1*/resources/*/files/*')
filter_ = [('xnat:mrSessionData/AGE', '>', '80'), 'AND']

for f in central.select(query).where(filter_):
    label = f.label()
    # images are stored in pairs of files (.img, .hdr) in this project
    if label.endswith('.img'):
        images.setdefault(label.split('.')[0], []).append(f)
    if f.label().endswith('.hdr'):
        images.setdefault(label.split('.')[0], []).append(f)
    # download and process both occur in parallel within the workers
    for name in images.keys():  
        if len(images[name]) == 2: # if .img and .hdr XNAT references are ready
            img, hdr = images.pop(name)                        # get references
            pool.apply_async(bet, (img, hdr), callback=notify) # start worker
pool.close()
pool.join()

\end{lstlisting}
\normalsize

\twocolumn

\end{document}